\documentclass[twocolumn,preprintnumbers,amsmath,amssymb,floatfix,prl]{revtex4}
\usepackage{amsmath}
\usepackage{dcolumn}
\usepackage{braket}
\usepackage{graphicx}
\usepackage{hyperref}
\begin{document}
\title{Physical Implementation of a Majorana Fermion Surface Code for \\ Fault-Tolerant  Quantum Computation} 
\author{Sagar Vijay and Liang Fu}
\affiliation{Massachusetts Institute of Technology Department of Physics\\ 77 Massachusetts Ave., Cambridge, MA 02139}
\begin{abstract}
We propose a physical realization of a commuting Hamiltonian of interacting Majorana fermions realizing $Z_{2}$ topological order, using an array of Josephson-coupled topological superconductor islands. The required multi-body interaction Hamiltonian is naturally generated by a combination of charging energy induced quantum phase-slips on the superconducting islands and electron tunneling. Our setup improves on a recent proposal for implementing a Majorana fermion surface code \cite{Maj_Surf_Code}, a `hybrid' approach to fault-tolerant quantum computation that combines (1) the engineering of a stabilizer Hamiltonian with a topologically ordered ground state with (2) projective stabilizer measurements to implement error correction and a universal set of logical gates. Our hybrid strategy has advantages over the traditional surface code architecture in error suppression and single-step stabilizer measurements, and is widely applicable to implementing stabilizer codes for quantum computation.  
\end{abstract}

\maketitle

\section{Introduction} 

Superconducting qubits based on Josephson junctions provide a promising platform for the implementation of a large-scale, fault-tolerant quantum computer \cite{Girvin}. 
The key physical degree of freedom that defines a superconducting qubit is 
the phase difference of the order parameter across a Josephson junction. 
In recent years, remarkable progress has been made on coherence times, high-fidelity gate operations and read-out protocols for superconducting qubits \cite{devoret, martinis, ibm}.

In addition to having a phase degree of freedom, superconductors host fermionic quasiparticles.  
In particular, quasiparticles with energies inside the superconducting gap can be localized in superconducting weak links or vortices. 
The quasiparticle occupation number could be exploited as an internal degree of 
freedom to encode information in a Fock space, opening up the possibility of 
{\it fermionic} quantum computation.  Of particular interest are Majorana fermions, excitations which arise as a spatially-localized ``half" of a zero-energy fermionic quasiparticle in a topological superconductor \cite{kitaev, Read_Green}. 

Several important considerations motivate us to   
study schemes for robust quantum computation using fermionic quasiparticles in superconductors.  From a quantum information perspective, it has been theoretically shown that  fermions have more computational power than bosons \cite{Bravyi_Kitaev}. 
Furthermore, a fermionic quantum computer can directly and efficiently simulate many-body Fermi systems with local interactions, 
without the computational cost of mapping to non-local bosonic systems.  

In this work, we propose a physical implementation of a quantum error-correcting code using an array of Majorana fermions as the underlying physical degrees of freedom, 
which was recently introduced and termed a Majorana fermion surface code \cite{Maj_Surf_Code}. Unlike the conventional surface code with bosonic physical qubits \cite{Surf_Code_I, Surf_Code_II, Surf_Code_III, Surf_Code_IV, Surf_Code_V}, the Majorana fermion surface code is based on commuting Hamiltonians of interacting Majorana fermions in two dimensions that exhibit $Z_2$ topological order \cite{Maj_Surf_Code, Xu_Fu, Terhal, Nussinov}. 
Importantly, our proposal combines the
{engineering} of the static parent Hamiltonian with projective measurements for active error correction and universal gate implementation. Our Hamiltonian-measurement
hybrid strategy has significant advantages over a measurement-only approach as employed in the conventional surface code, 
and provides a general approach to implementing stabilizer codes.

Our hybrid strategy is particularly adapted to providing the error correction capability that is {required} for {a} {scalable} quantum computing architecture.  For \emph{any} quantum computer operating at non-zero temperature, a {finite density} of spurious excitations will be thermally excited, interfering with encoded information 
and leading to the failure of fault-tolerance in the absence of active error correction.  As a result,  Hamiltonian-only approaches to quantum computation, including those based on non-Abelian topological phases, 
will be unable to proceed correctly in the absence of measurement-based error correction.  

In contrast, our Hamiltonian-measurement hybrid approach achieves scalability and fault-tolerance in two ways. First, engineering an ideal stabilizer Hamiltonian naturally provides error suppression at temperatures below the energy gap, and remarkably permits {\it single-step} measurements of {\it multi-body} stabilizer operators, which minimizes readout errors. Furthermore, as the interactions between the physical degrees of freedom are described by the stabilizer Hamiltonian in our setup, we are able to take non-idealities in stabilizer measurements into account in a controlled fashion.  Second, the constant projective measurements of stabilizers can detect unwanted excitations (i.e., stabilizer flips) that do occur.  By collecting measurements at several time-steps, an error can be optimally decoded and used to correct a logical qubit during readout as in the conventional surface code, leading to a scalable quantum computing architecture \cite{Surf_Code_V, Surf_Code_Error}.

This paper is organized as follows. First, we introduce a commuting Hamiltonian  of interacting Majorana fermions that exhibits a $Z_{2}$ topological order of Fermi systems, 
as thoroughly described in \cite{Maj_Surf_Code}.  Next, we demonstrate that this exact Hamiltonian can be realized in an array of Josephson-coupled topological superconductors, and  
 projective measurements of the commuting stabilizers can be achieved using current superconducting qubit technology. This combination of engineering a static Hamiltonian and performing constant measurements constitutes a full implementation of a Majorana fermion surface code. Potential advantages 
 of our implementation over previous proposals \cite{Maj_Surf_Code, Terhal} are discussed.  
 For the sake of completeness, we also briefly review how logical qubits may be encoded and manipulated to achieve a universal set of gates, as detailed in \cite{Maj_Surf_Code}. 
\section{Fermionic $Z_{2}$ Topological Order}

\begin{figure}
\includegraphics[trim = 0 74 164 0, clip = true, width=0.37\textwidth, angle = 0.]{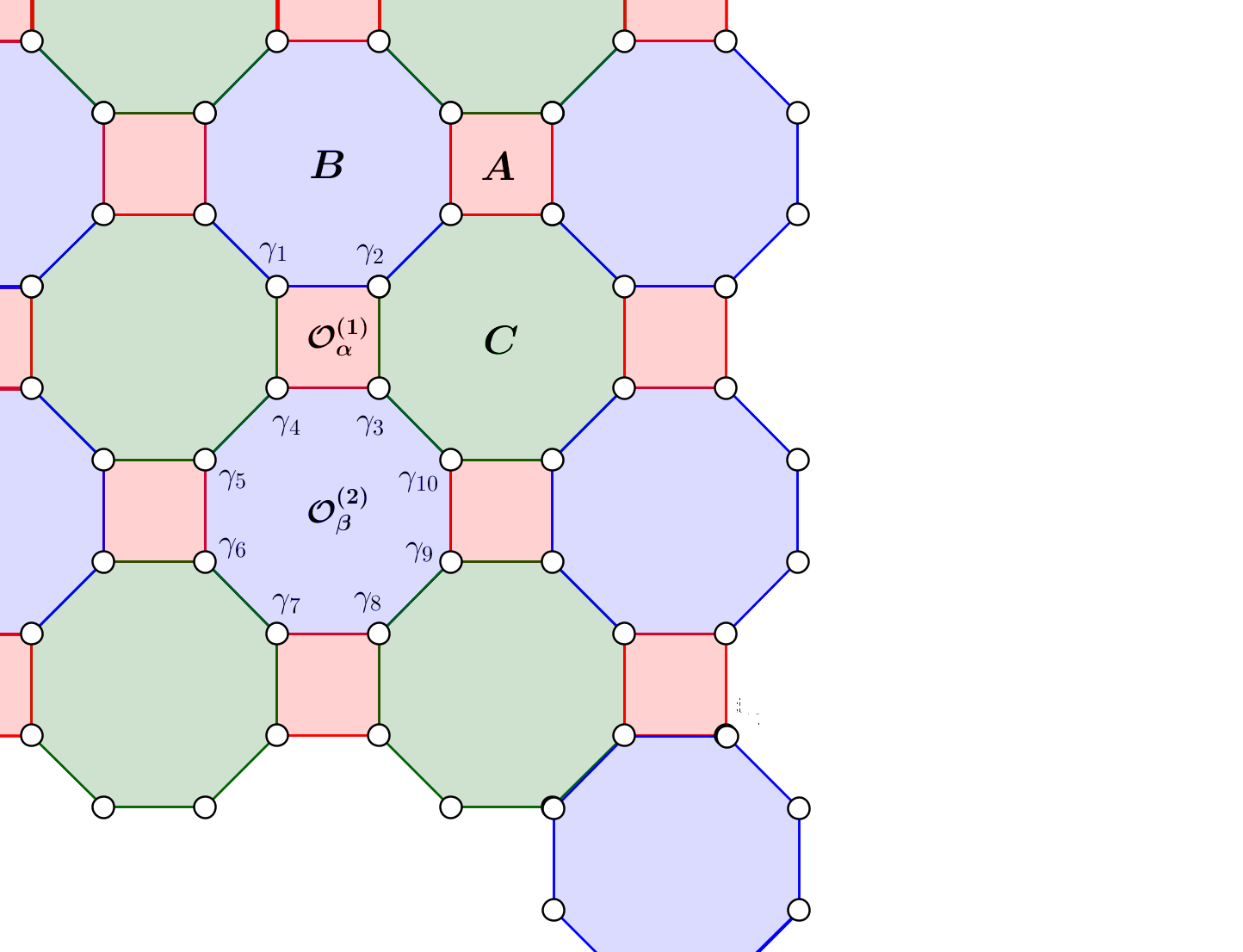} 
\caption{We consider a square/octagon lattice with one Majorana fermion per site as shown.  The Hamiltonian (\ref{eq:Hamiltonian}) is the sum of the fermion parity of each plaquette.  We label the plaquettes $A$, $B$, and $C$ (as indicated by the coloring), corresponding to the three distinct types of fundamental excitations that can be created by acting on the ground-state with string-like Wilson line operators.}\label{fig:Lattice}
\end{figure}

We begin by introducing a commuting Hamiltonian that realizes a fermionic $Z_{2}$ topological order. Consider a square-octagon lattice with one Majorana fermion per lattice site, as shown in Figure \ref{fig:Lattice}a.  The Hamiltonian is defined as
\begin{align} \label{eq:Hamiltonian}
H_{0} = - u_1 \sum_{\alpha}\mathcal{O}_{\alpha}^{(1)} - u_2 \sum_{\beta}\mathcal{O}_{\beta}^{(2)} 
\end{align}
where $\mathcal{O}_{\alpha}^{(1)}$ is the product of the four Majorana fermions bordering a square plaquette $\alpha$, while $\mathcal{O}_{\beta}^{(2)}$ is the product of the eight Majoranas around an octagonal plaquette $\beta$.  Any pair of Majorana fermions ($\gamma$) anti-commute and satisfy $\gamma^{2} = 1$. Therefore, each of the operators appearing in the Hamiltonian squares to the identity, and has eigenvalues $\pm 1$. Furthermore, as any pair of operators in the Hamiltonian overlap on an even number of Majorana fermions, all terms in the Hamiltonian mutually commute, and constitute a complete set of stabilizer operators. The ground state satisfies $\mathcal{O}_{\alpha}^{(1)}\ket{\Psi_{\mathrm{gs}}} = \ket{\Psi_{\mathrm{gs}}}$ and $\mathcal{O}_{\beta}^{(2)}\ket{\Psi_{\mathrm{gs}}} = \ket{\Psi_{\mathrm{gs}}}$ for all plaquettes $\alpha$, $\beta$.

The Hamiltonian (\ref{eq:Hamiltonian}) belongs to a family of solvable ``Majorana plaquette models'' in two dimensions \cite{Maj_Surf_Code, Bravyi}, and arises as a low-energy effective Hamiltonian studied in the work of Xu and Fu \cite{Xu_Fu} and others \cite{Terhal, Nussinov}.  
The ground state exhibits four-fold topological degeneracy on the torus, and realizes a $Z_{2}$ topological order of Fermi systems.  Topological excitations are created from the ground state by acting with the product of Majorana fermions along lines; this flips the eigenvalues of plaquette operators at the ends of the lines, creating a pair of topological excitations. The three fundamental excitations -- labeled $A$, $B$, or $C$ according to the plaquette type defined in Figure \ref{fig:Lattice} -- may only be created in pairs and cannot be deformed into each other.  These three types of excitations are bosons, but have $\pi$ mutual statistics relative to each other; see \cite{Maj_Surf_Code} for a detailed discussion.

We propose an exact physical realization of the Hamiltonian (\ref{eq:Hamiltonian}) using an array of square and octagonal islands of two-dimensional (2D) topological superconductors, which can be realized by coupling $s$-wave superconductors to topological insulator (TI) surface states \cite{Fu_Kane}, or spin-orbit-coupled 2D electron gases \cite{2DEG1, 2DEG2}. By applying magnetic flux or current bias, we fix the phases of the superconducting islands  in the configuration shown in Figure \ref{fig:Phase_Slip}a, so that 
the phases of any three neighboring islands wind by $\pm 2\pi$ around each tri-junction, leading to a 2D array of Josephson vortices.   
As demonstrated by Fu and Kane \cite{Fu_Kane}, a phase vortex or anti-vortex in the proximity-induced superconductivity on the TI surface binds a localized Majorana zero mode.  As a result, the phase configuration of the superconducting islands shown in Figure \ref{fig:Phase_Slip}a produces the desired square-octagon lattice of Majorana fermions. 

\section{Physical Realization of the Stabilizer Hamiltonian}
 
\begin{figure}
\includegraphics[trim = 0 70 150 0, clip = true, width=0.37\textwidth, angle = 0.]{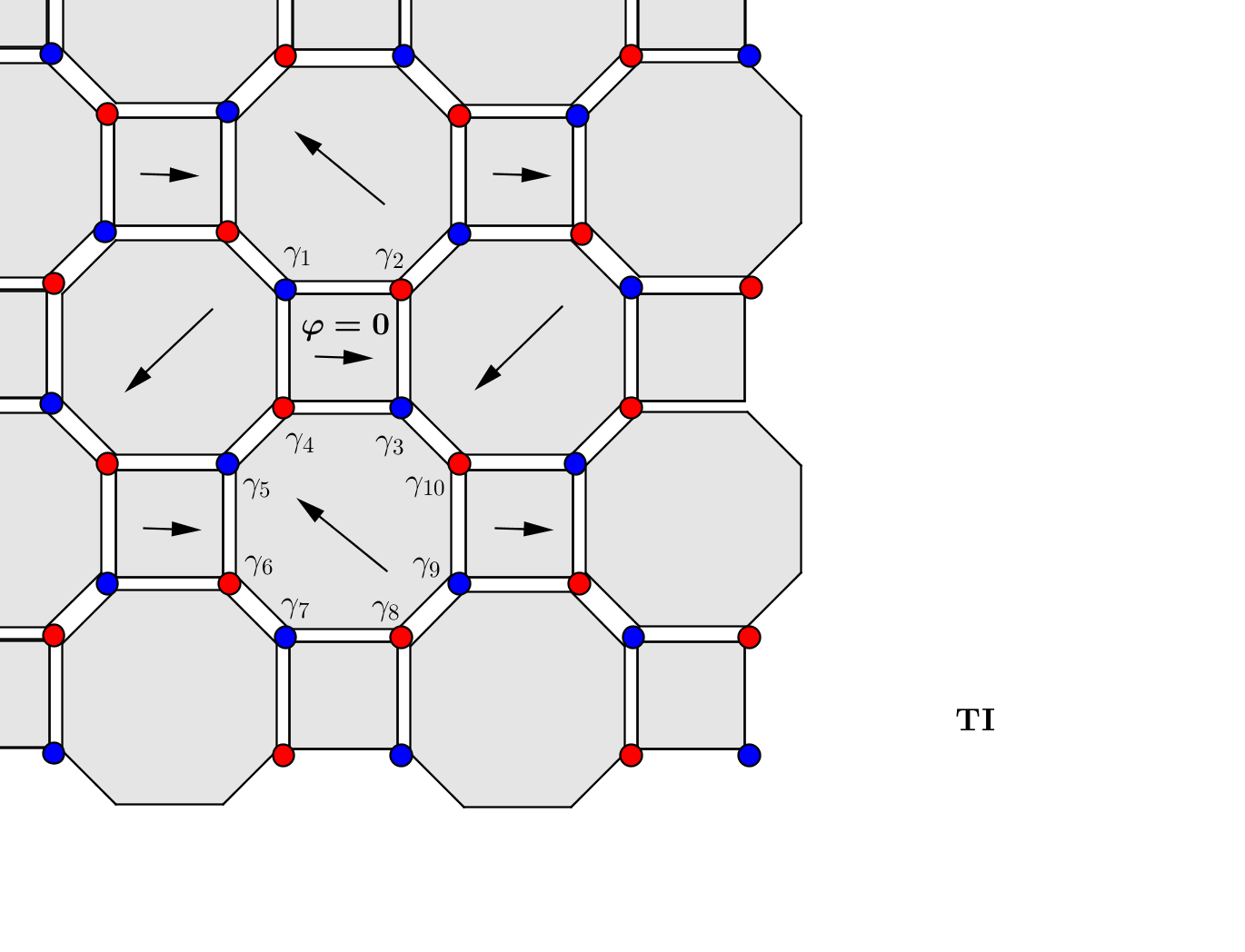}
\caption{We place an array of $s$-wave superconductors (grey) on a topological insulator surface, with the arrows indicating the relative phase of each superconducting island.  The phase winds by $\pm 2\pi$ around each tri-junction and traps a Majorana zero mode, as originally demonstrated in \cite{Fu_Kane}.  Quantum phase-slips on the square islands generate a four-Majorana interaction that couples the zero modes at the adjacent tri-junctions.  By making the size of the octagonal islands sufficiently small, the wavefunction overlap of the zero modes generates an eight-Majorana interaction (e.g. $\mathcal{O}_{\beta}^{(2)} = \gamma_{4}\gamma_{5}\gamma_{6}\gamma_{7}\gamma_{8}\gamma_{9}\gamma_{10}\gamma_{3}$) at each octagonal island.}\label{fig:Phase_Slip}
\end{figure}

We now engineer the 4-body and 8-body interactions in the Hamiltonian (\ref{eq:Hamiltonian}). First, we show that the 4-body interaction is naturally generated by the charging energy on the square superconducting islands.  Our analysis closely follows Ref. \cite{Maj_Surf_Code}, where it was first demonstrated that the effect of superconducting phase-fluctuations on Majorana zero modes bound to vortex cores may be used to engineer non-local, multi-fermion interactions. Let us  consider the effect of a small charging energy $E_c$ on a single square island $\alpha$, holding the phases of all other islands fixed. The Hamiltonian for this island is given by 
\begin{align}\label{eq:H_j}
H_{\alpha}(n_{g}) = \,\, &E_{c} \left(- i \frac{\partial}{\partial \varphi_\alpha} - n_{g} \right)^{2}\nonumber\\
&- E_{J}\sum_{\langle \alpha,\,\beta\rangle}\cos(\varphi_{\alpha} - \varphi_\beta - a_{\alpha\beta}).
\end{align}
The first term describes the Coulomb energy due to the excess charge on an island, which is capacitively coupled to ground and to the other islands in the array. Here $n_{g}$ is an offset charge that can be continuously tuned by a gate voltage.  The second term describes the Josephson coupling of neighboring islands with energy $E_{J}$, with $a_{jj'}$ tuned so as to produce the superconducting phase configuration in Figure \ref{fig:Phase_Slip}a. Importantly,  as the charging energy and the Josephson energy do not commute, the superconducting phase $\varphi_\alpha$ is a quantum-mechanical variable. 

\begin{figure*}
$\begin{array}{cccc}
\includegraphics[trim = 15 67 163 0, clip = true, width=0.25\textwidth, angle = 0.]{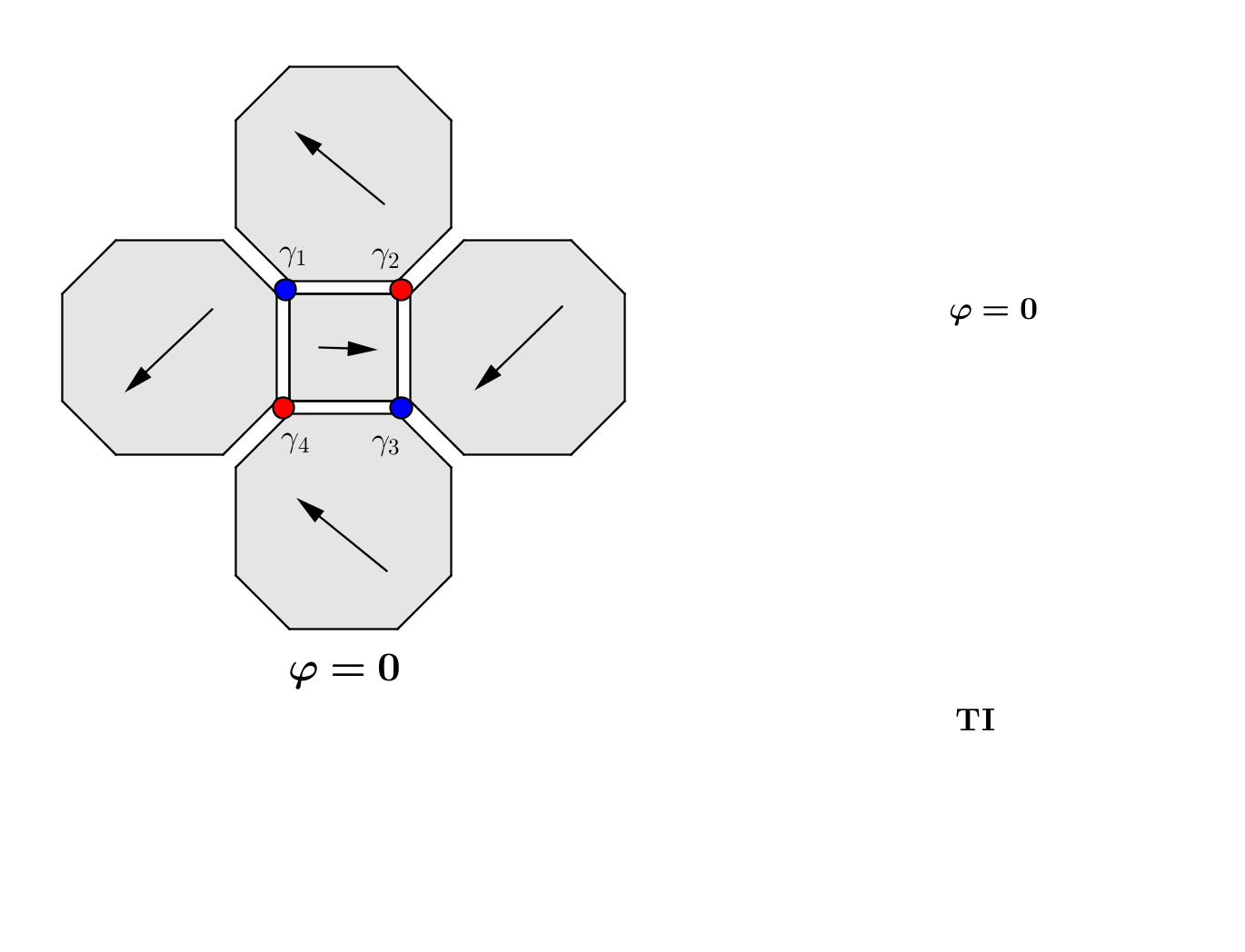} &
\includegraphics[trim = 15 67 163 0, clip = true, width=0.25\textwidth, angle = 0.]{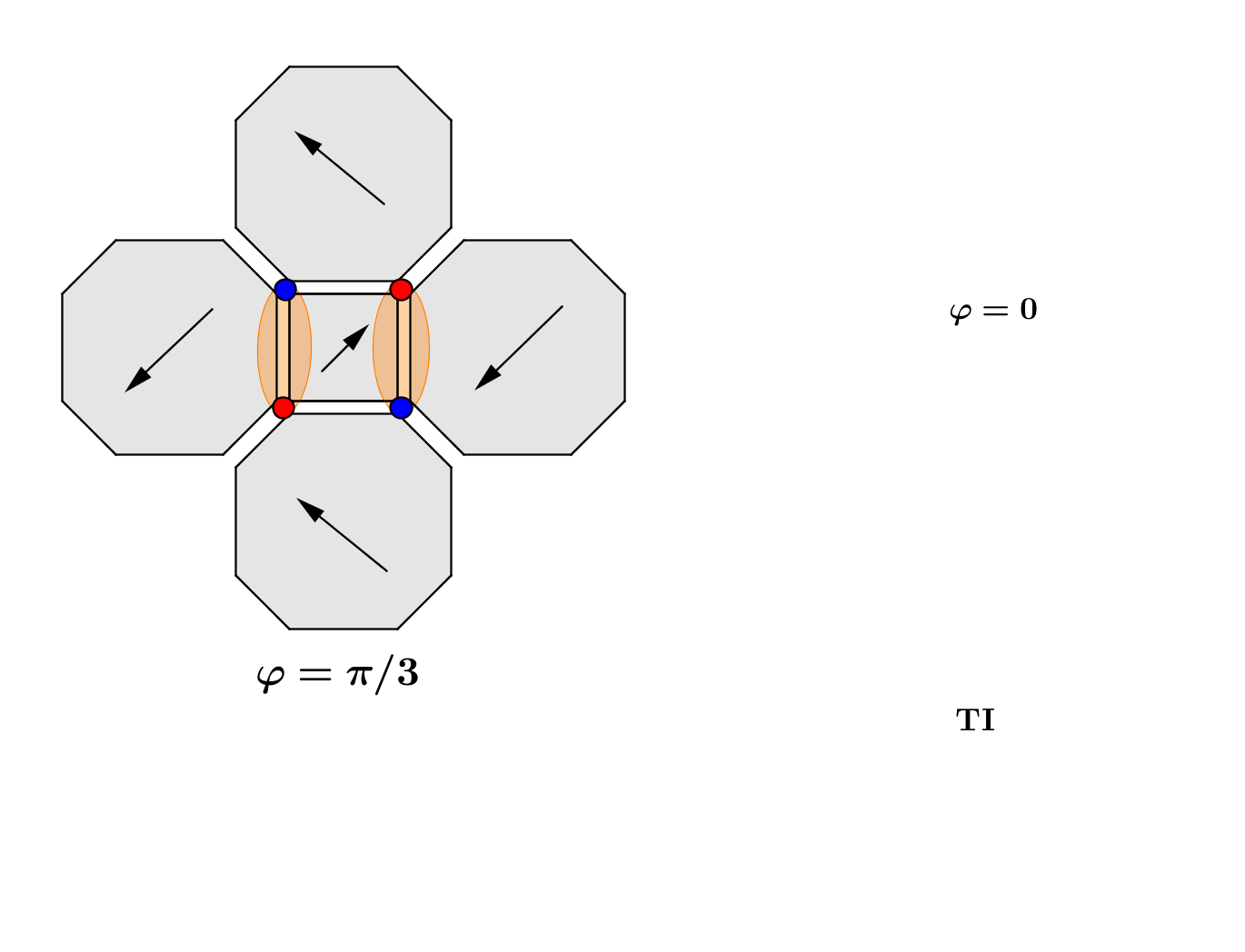} &
\includegraphics[trim = 15 67 163 0, clip = true, width=0.25\textwidth, angle = 0.]{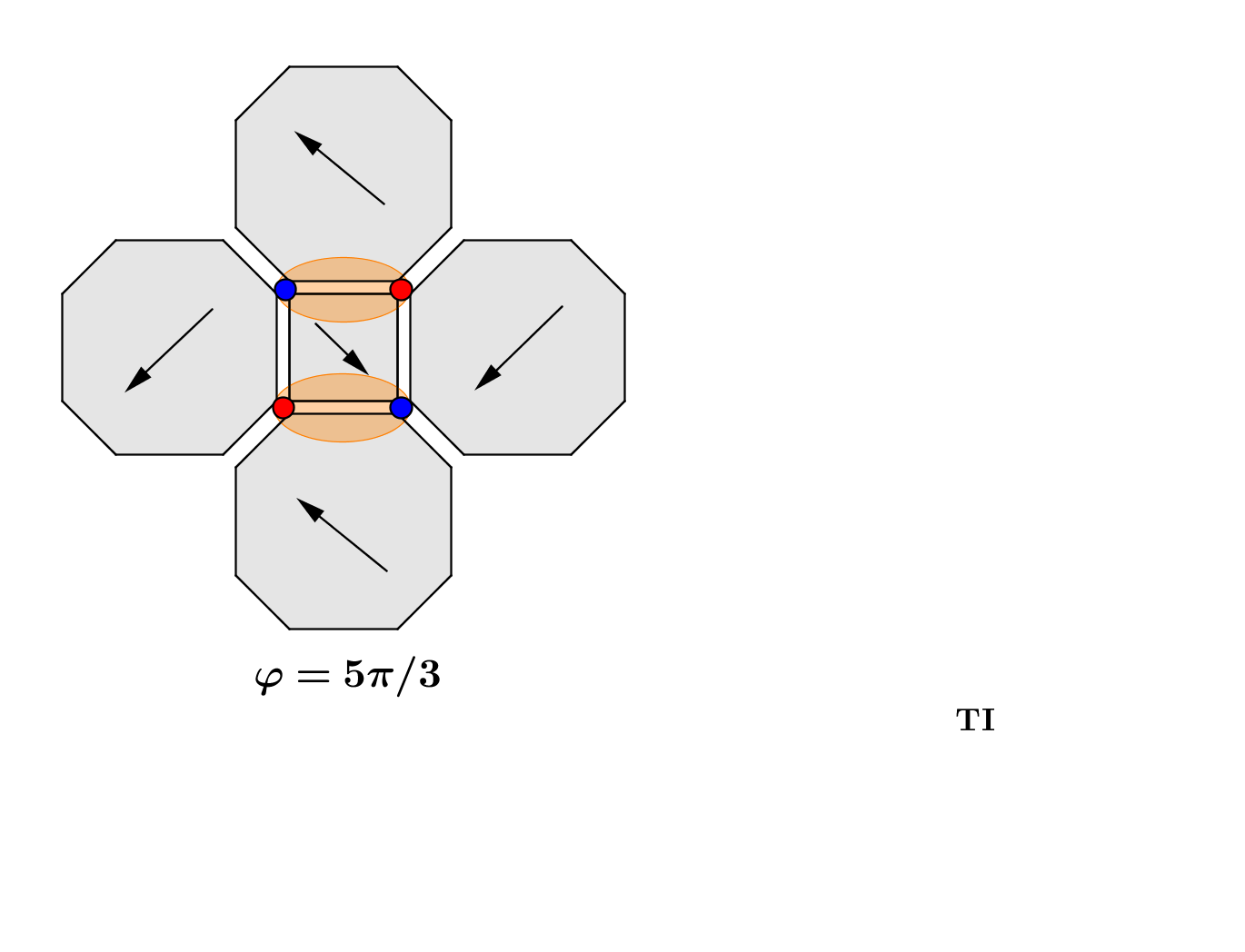} &
\includegraphics[trim = 15 67 163 0, clip = true, width=0.25\textwidth, angle = 0.]{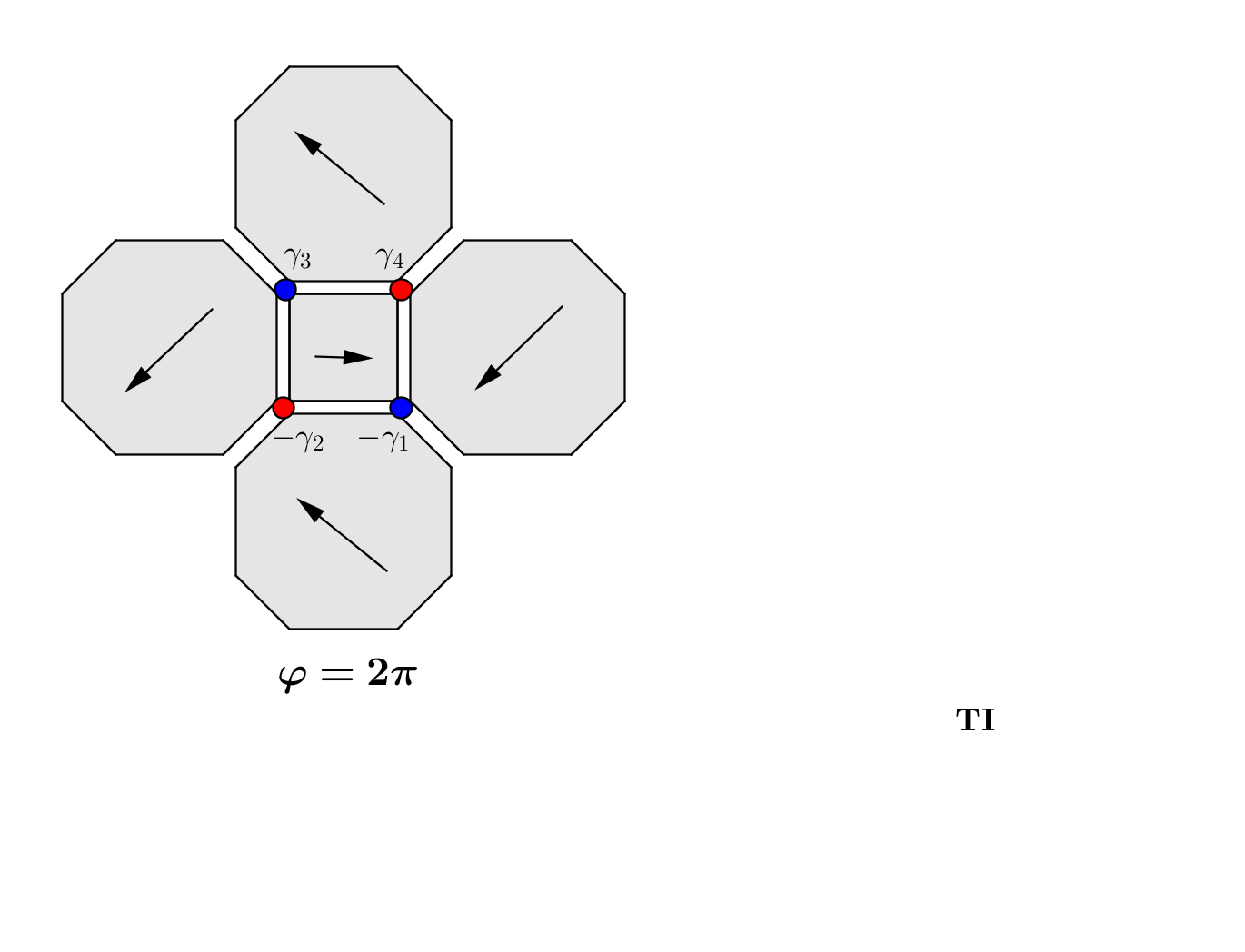}\\ \\
\text{(a)} & \text{(b)} & \text{(c)} & \text{(d)}
\end{array}$
\caption{A $+2\pi$ phase slip on a square superconducting island exchanges the Majorana zero modes bound at vortices (red) as well as the zero modes at anti-vortices (blue).  The panels (a)-(d) show the winding of the superconducting phase $\varphi$ of the square island; when the phase of the square island is anti-aligned with the phase of adjacent islands, pairs of Majorana zero modes along their common boundary strongly couple, as shown in (b) and (c), and are exchanged. The full $+2\pi$ phase-slip process results in the transformation $\gamma_{1}\rightarrow\gamma_{3}, \gamma_{3}\rightarrow-\gamma_{1}, \gamma_{2}\rightarrow\gamma_{4}, \gamma_{4}\rightarrow-\gamma_{2}$ as shown in (d). }\label{fig:Phase_Slip_Mechanism}
\end{figure*}

A small charging energy $E_{c} \ll E_{J}$ induces quantum fluctuations of the superconducting phase on the square island.  Small fluctuations result in a harmonic oscillator spectrum with level-spacing $\epsilon_{0} \sim \sqrt{E_{J}E_{c}}$.  As originally found in \cite{Maj_Surf_Code}, a remarkable feature of our setup is that \emph{quantum phase slips} $\varphi_{\alpha} \rightarrow \varphi_\alpha + 2\pi m$ permute vortices at tri-junctions and hence the Majorana fermions bound to these vortices. 
Specifically, a $2\pi$ phase slip on a square island  exchanges the two Majorana fermions bound to the (anti-)vortices at diagonally opposite vertices in a (counter-)clockwise manner, via a sequence of 
intermediate configurations shown in Figures \ref{fig:Phase_Slip_Mechanism}a-\ref{fig:Phase_Slip_Mechanism}d. The net effect of this phase slip is to {\it permute} the Majorana fermions as follows
\begin{align}
\gamma_{1}  \rightarrow \gamma_{3} ,    \gamma_{3}  \rightarrow  -\gamma_{1}, \nonumber \\
\gamma_{2}  \rightarrow  \gamma_{4} , \gamma_{4}  \rightarrow  -\gamma_{2}.\label{transform}
\end{align}
The braiding of Majorana fermions in charing-energy induced quantum phase slip processes is crucial to our implementation of the Majorana fermion surface code and enables a new way of detecting the non-Abelian statistics of Majorana zero modes \cite{forthcoming}.    

As shown in \cite{Maj_Surf_Code},  the low-energy effective Hamiltonian for the island in the limit $E_{c} \ll E_{J}$ takes the form of a tight-binding Hamiltonian describing a ``phase'' particle that tunnels between the minima of the periodic Josephson potential, located at $\varphi_\alpha + 2\pi m$. For a $2\pi$ phase-slip, the contribution to the effective Hamiltonian is given by
 \begin{align}
H_{\alpha}(n_{g}) = \epsilon_{0} + \left(  t_\alpha \; \hat{U} e^{2\pi i n_{g}} + \mathrm{h.c.} \right).  \label{heff}
\end{align}
Here $\epsilon_0$ is the on-site energy of the lowest harmonic oscillator state associated with each potential minimum, while
$t_\alpha$ is the tunneling amplitude between nearest neighbors, given by the amplitude of a $2\pi$ phase slip $t_\alpha \propto e^{-\sqrt{2E_J/E_c}}$. 
The offset charge $n_g$ produces a Berry phase $e^{2\pi i n_{g}}$ in the effective Hamiltonian. Importantly,  
 in our setup the phase particle carries an internal $2^{4/2}=4$-dimensional Hilbert space resulting from the four Majorana fermions at the vertices.   
Here $\hat{U}$ is the unitary operator that implements the braiding of Majorana fermions in a $2\pi$ phase slip process, as described in (\ref{transform}), and is given by   
\begin{align}
\hat{U} = \frac{1 + \gamma_{1}\gamma_{3}}{\sqrt{2}}\cdot\frac{1 + \gamma_{2}\gamma_{4}}{\sqrt{2}}. \label{U}
\end{align}
Substituting (\ref{U}) into (\ref{heff}), we obtain the following effective Hamiltonian for the lowest harmonic oscillator level on a square island $\alpha$:
\begin{align}\label{eq:H_eff}
H_{\alpha}(n_{g}) & =  \epsilon_{0}  - t_\alpha \cos (2\pi  n_g ) \gamma_1 \gamma_2 \gamma_3 \gamma_4  \nonumber \\
&+  t_\alpha \sin (2\pi  n_g ) (i\gamma_{1}\gamma_{3} + i\gamma_{2}\gamma_{4}),   
\end{align}
where we have eliminated the $U(1)$ phase factor of $t_{\alpha}$ by shifting $n_g$ by a constant. 
From now on, we set $2n_g$ to integer values by tuning the gate voltage, so that $H_{\alpha}$ is exactly the $4$-body interaction on square plaquettes needed for the Majorana plaquette model (\ref{eq:Hamiltonian}). We further require the size of each square island to be sufficiently large so that  the four Majorana fermions 
at its corners  have negligible wavefunction overlap, thus preventing any unwanted bilinear coupling within square islands.


\begin{figure*}
$\begin{array}{cccc}
\includegraphics[trim = 4 68 173 4, clip = true, width=0.24\textwidth, angle = 0.]{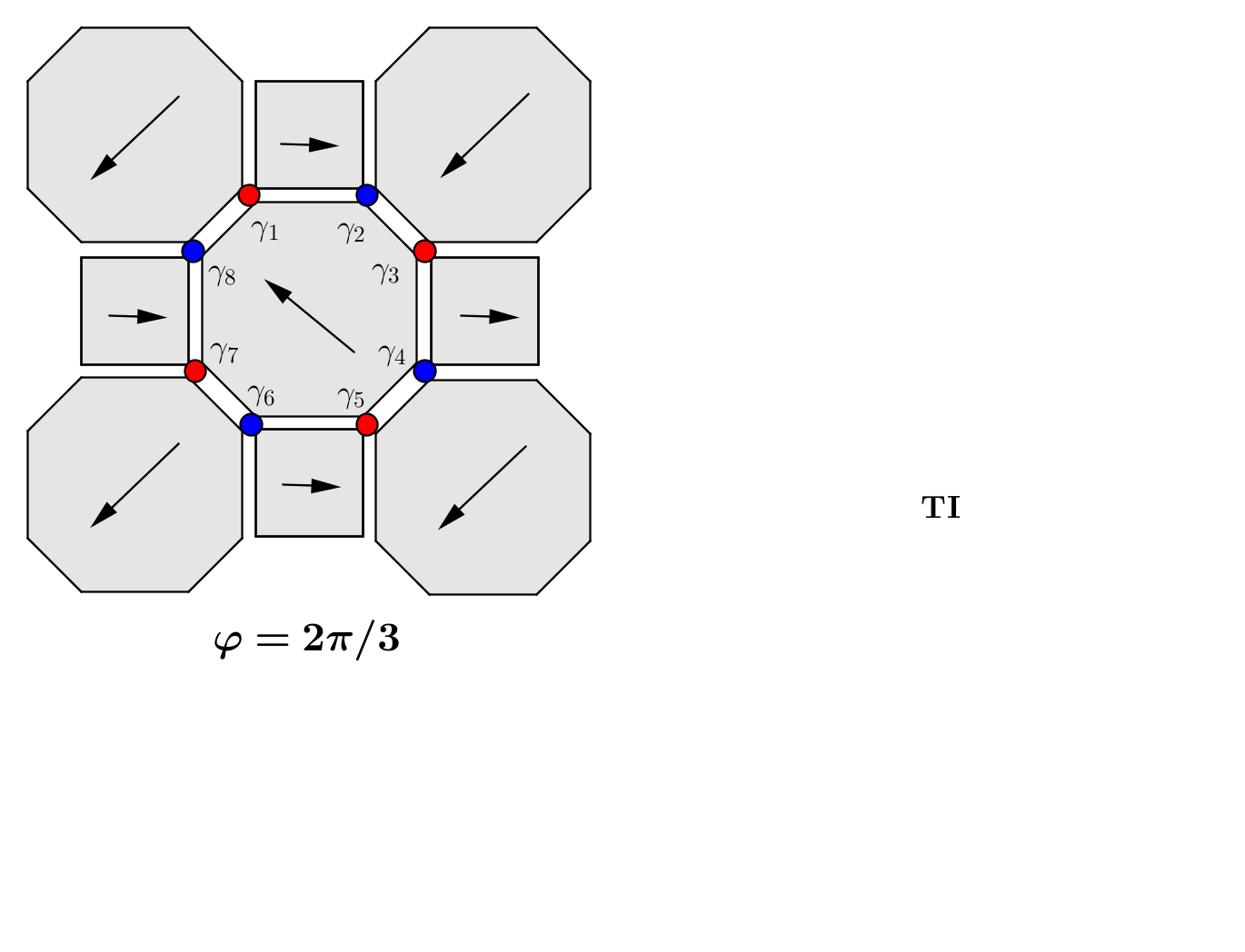} &
\includegraphics[trim = 4 68 173 4, clip = true, width=0.24\textwidth, angle = 0.]{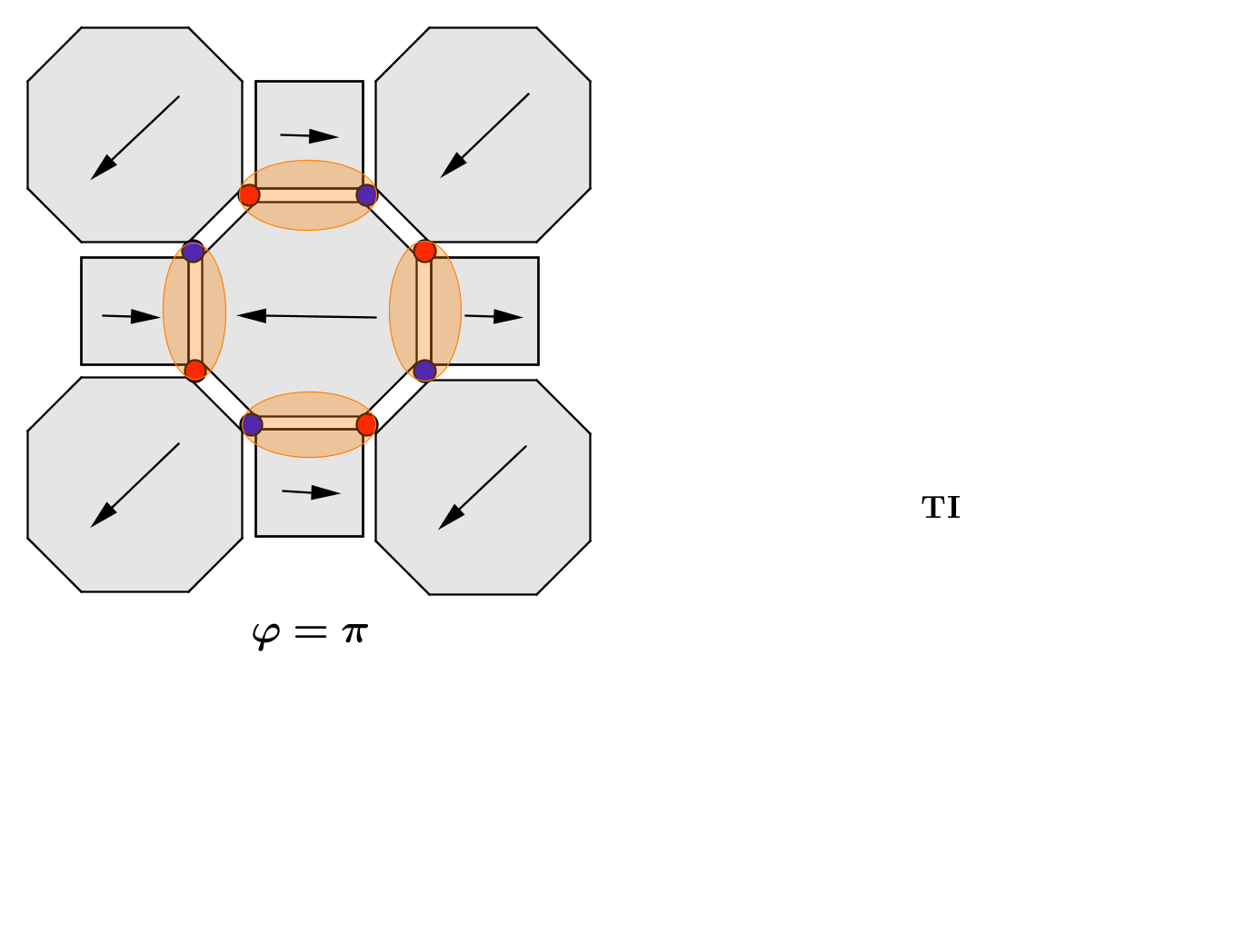} &
\includegraphics[trim = 4 68 173 4, clip = true, width=0.24\textwidth, angle = 0.]{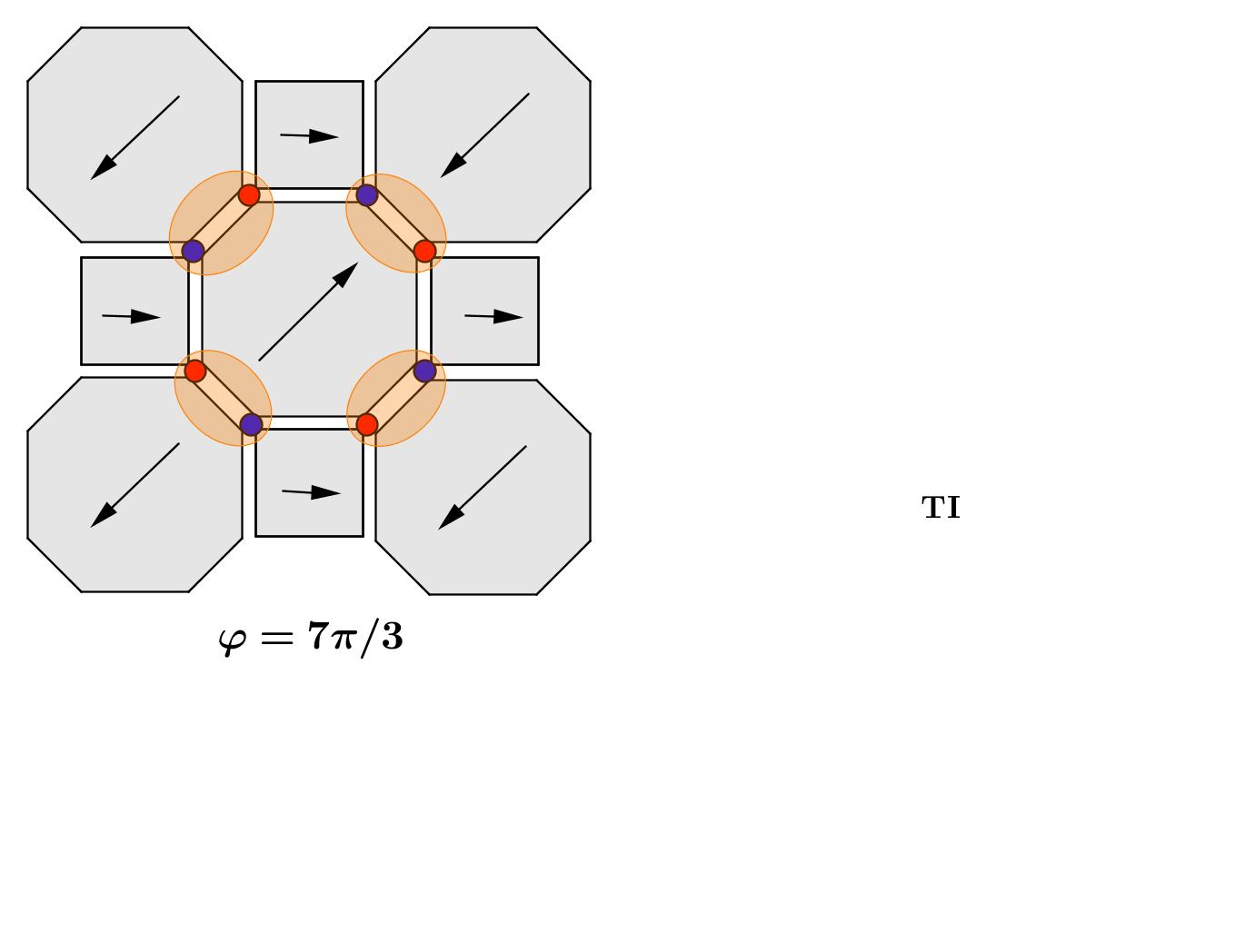} &
\includegraphics[trim = 4 67 173 4, clip = true, width=0.24\textwidth, angle = 0.]{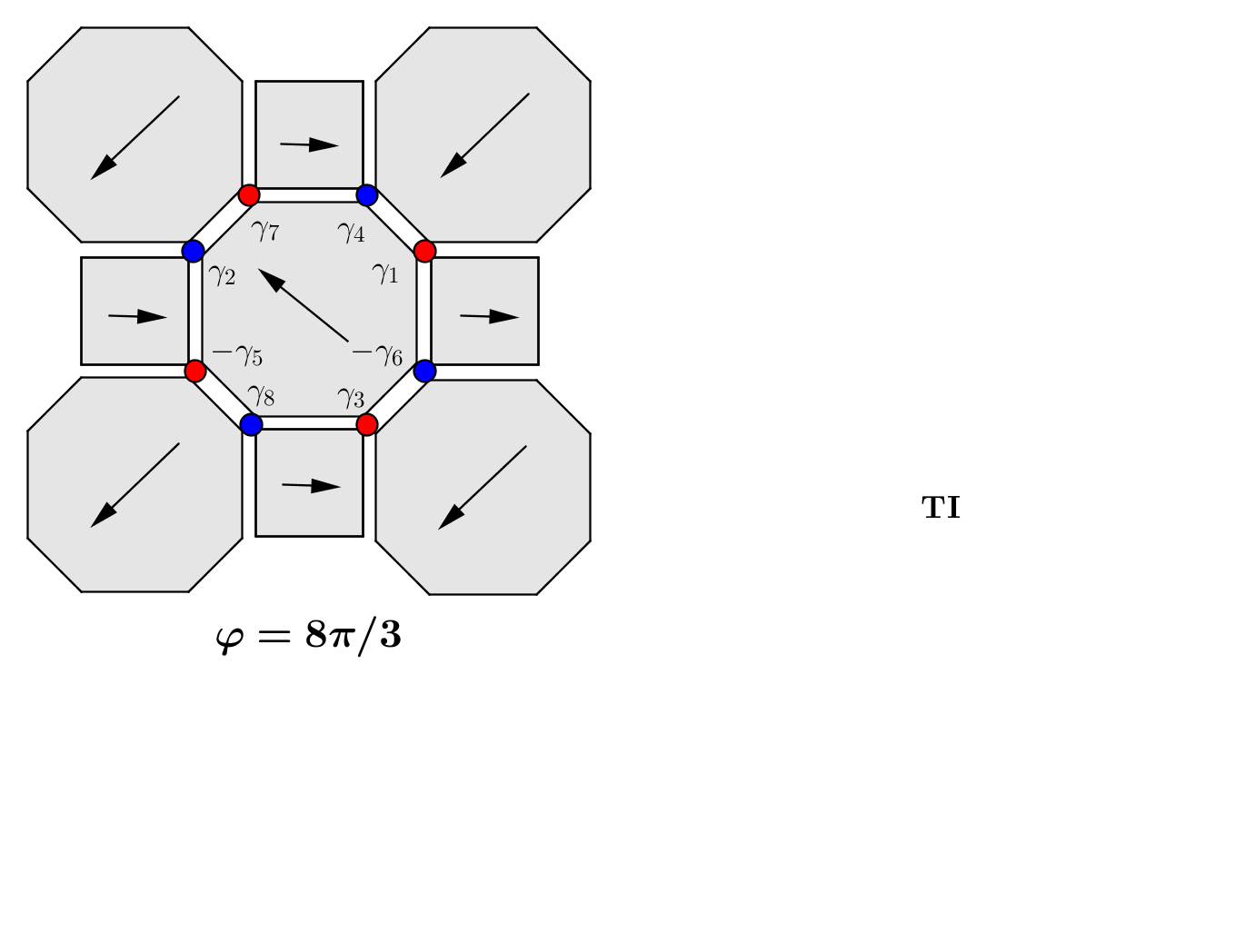}\\
\text{(a)} & \text{(b)} & \text{(c)} & \text{(d)}
\end{array}$
\caption{A $+2\pi$ phase slip on an octagonal superconducting island braids the Majorana zero modes bound at vortices (red) as well as the zero modes at anti-vortices (blue) as shown.  The panels (a)-(d) show the winding of the superconducting phase $\varphi$ of the octagonal island. The full $+2\pi$ phase-slip process results in the transformation $\gamma_{1}\rightarrow\gamma_{7}$, $\gamma_{3}\rightarrow\gamma_{1}$, $\gamma_{5}\rightarrow\gamma_{3}$, $\gamma_{7}\rightarrow-\gamma_{5}$,
$\gamma_{2}\rightarrow\gamma_{4}$, $\gamma_{4}\rightarrow-\gamma_{6}$, $\gamma_{6}\rightarrow\gamma_{8}$, $\gamma_{8}\rightarrow\gamma_{2}$ as shown in (d). }\label{fig:Phase_Slip_Mechanism_2}
\end{figure*}

To further implement the 8-body interaction on the octagonal islands, we consider the electron tunneling between adjacent square islands 
via nearest-neighbor Majorana fermion zero modes at tri-junctions. Such tunneling processes are described by Majorana bilinear operators such as $i \delta \gamma_4 \gamma_5$ in Figure \ref{fig:Phase_Slip}. 
The tunneling amplitude $\delta$ depends exponentially on the the wavefunction overlap between $\gamma_4$ and $\gamma_5$, 
which can be tuned by adjusting the distance between adjacent square islands and other device parameters (e.g., the chemical potential). 
We work in the regime $\delta \ll t_\alpha$, and treat the effect of $\delta$ perturbatively. 
A single-electron tunneling event flips the fermion parity of two adjacent square islands and leads to an excited state with a large energy cost of $2 t_\alpha$.    
The lowest order process that 
brings the system back to the ground state manifold of $H_\alpha$ consists of four single-electron tunneling events between four pairs of nearest-neighbor Majorana fermions, 
such as $(\gamma_4, \gamma_5), (\gamma_6, \gamma_7), (\gamma_8, \gamma_9), (\gamma_{10}, \gamma_3)$ shown in Figure \ref{fig:Phase_Slip}.  
From fourth-order perturbation theory, the effective Hamiltonian for an octagonal island  is found to be a ring-exchange term: 
\begin{align}\label{eq:H_Oct}
H_{\beta} = -\frac{5\delta^{4}}{16 t_\alpha^{3}}\mathcal{O}_{\beta}^{(2)}
\end{align}
where $\mathcal{O}_{\beta}^{(2)}$ is the product of the 8 Majorana fermions on the vertices of an octagonal island $\beta$. 
The sum of the 4-body interaction (\ref{eq:H_eff}) and the 8-body interaction (\ref{eq:H_Oct}) in our setup of the topological superconductor array  
{\it precisely} yields the commuting Hamiltonian (\ref{eq:Hamiltonian}), whose ground states and excitations 
will hereafter be used to encode quantum information, leading to a Majorana fermion surface code.

Physical realizations of Majorana plaquette models have been previously proposed in other platforms for Majorana fermions. 
In Ref. \cite{Xu_Fu}, Xu and Fu 
introduced a square-octagon lattice of Majorana fermions in an array of quantum spin Hall insulators and superconductors, and 
derived the Hamiltonian (\ref{eq:Hamiltonian}) when charging energy on the square superconducting islands and electron tunneling between the islands are present.   
The same approach was later employed to implement (\ref{eq:Hamiltonian}) in a network of semiconductor nanowires \cite{Terhal}. 
In these works, Majorana fermions sit at the interface between superconducting and insulating regions, rather than being bound to vortices as in our proposal---a 
crucial feature for implementing the stabilizer measurements in our Majorana fermion surface code, as discussed below.    
In our recent work with Hsieh \cite{Maj_Surf_Code}, a physical realization of a Majorana plaquette model  in an array of hexagonal topological superconductor islands was studied. 
However, it was found that charging energy generates unwanted two-body interactions, in addition to the desired 6-body interactions between Majorana fermions. 
As we have shown above, this work improves on Ref. \cite{Maj_Surf_Code} and achieves an exact realization of the Majorana plaquette model (\ref{eq:Hamiltonian}) in an array of 2D topological superconductor islands.

\section{Measurement Protocol and Universal Quantum Computation}

As proposed in Ref. \cite{Maj_Surf_Code}, we may use this physical realization of a fermion model with $Z_2$ topological order as a platform for universal quantum computation -- a Majorana fermion surface code.  The ground state of the Hamiltonian $H_0$ is a highly-entangled many-body state and serves as a reference code state. 
Thermal excitations at non-zero temperature are stabilizer flips, and can only be created in pairs on the same types of plaquettes.  
However, even at low temperatures, the number of flips is given by $N_{t} \propto N_0 e^{-u/ k_B T}$, where $u\equiv 
  \min(u_1, u_2)$ is the minimum energy cost for a stabilizer flip and $N_0$ is the total number of stabilizers in the system.  
Therefore, despite being suppressed by the Boltzmann factor, the number of stabilizer flips scales extensively with the system size at a fixed temperature, so that any reliable, large-scale quantum computation necessarily requires active quantum error correction. 
To detect and handle errors in the Majorana fermion surface code, 
we perform constant measurements of all commuting stabilizers, except those used to encode logical qubits (see below), to    
project the system onto a stabilizer eigenstate.  The measured stabilizer eigenvalues are recorded and used to correct errors, as detailed in  Ref. \cite{Maj_Surf_Code} 
and references therein.

Logical qubits are encoded by stopping the measurement of two stabilizers of the same type ($A$, $B$ or $C$) in subsequent cycles, so that the presence or absence of an anyon 
(= stabilizer flip) in the resulting ``hole" defines the two states of the encoded qubit.  A set of gates required for universal quantum computation (CNOT, Hadamard, $S$ and $T$) may be implemented by performing a sequence of measurements that have the effect of moving the logical qubits.  A detailed description of gate and measurement protocols for the Majorana surface code is provided in \cite{Maj_Surf_Code}.  Notably, since lattice symmetries exchange the $B$ and $C$ anyons, the Hadamard gate has a much simpler implementation than in the ordinary surface code \cite{Maj_Surf_Code}.

We now describe how to perform projective measurements of the commuting stabilizers in our setup. As discussed in \cite{Maj_Surf_Code}, the eigenvalue of a four-Majorana stabilizer $\mathcal{O}_\alpha^{(1)}$ can be determined by exciting the square island $\alpha$ with microwave photons and measuring the energy gap to the next harmonic oscillator level  \cite{Hassler, Glazman}.  As the sign of $u_1$ associated with the stabilizer operator in the Hamiltonian $H_0$ alternates between consecutive harmonic oscillator levels, the gap depends sensitively on the stabilizer eigenvalue. Therefore, a measurement of the energy gap directly determines $\mathcal{O}_{\alpha}^{(1)}$.

Our implementation of the eight-Majorana stabilizer measurement deserves special attention. 
To measure the eigenvalue of $\mathcal{O}_{\beta}^{(2)}$, we increase the charging energy on an octagonal superconducting island 
to activate quantum phase-slips. Importantly, we require that the charging energy is small relative to the energy cost for flipping any stabilizer eigenvalue, 
and the charging energy is increased sufficiently slowly, on a time-scale larger than the inverse gap $\tau \gg t_{\alpha}^{3}/\delta^{4} \gg 1/t_\alpha$. 
Under this adiabatic condition, transitions to excited stabilizer eigenstates are avoided.

As in the case of square islands,  a small charging energy on an octagon island induces quantum phase slips that permutes the eight Majorana zero modes at the vertices. 
As shown in Figure \ref{fig:Phase_Slip_Mechanism_2}a-\ref{fig:Phase_Slip_Mechanism_2}d, a $2\pi$ phase slip has the effect of permuting the Majorana zero modes as follows:
\begin{align}
&\gamma_{1}\rightarrow\gamma_{7},\,\,\gamma_{3}\rightarrow\gamma_{1},\,\,\,\gamma_{5}\rightarrow\gamma_{3},\,\,\,\gamma_{7}\rightarrow-\gamma_{5}\\
&\gamma_{2}\rightarrow\gamma_{4},\,\,\gamma_{4}\rightarrow-\gamma_{6},\,\,\,\gamma_{6}\rightarrow\gamma_{8},\,\,\,\gamma_{8}\rightarrow\gamma_{2}.
\end{align}
This transformation is implemented by the unitary operator $\hat{W} = \mathcal{O}_{1}\mathcal{O}_{2}$ where
\begin{align}
&\mathcal{O}_{1} \equiv \frac{1 + \gamma_{1}\gamma_{2}}{\sqrt{2}}\frac{1 + \gamma_{3}\gamma_{4}}{\sqrt{2}}\frac{1 - \gamma_{5}\gamma_{6}}{\sqrt{2}}\frac{1 + \gamma_{7}\gamma_{8}}{\sqrt{2}}\\
&\mathcal{O}_{2} \equiv \frac{1 + \gamma_{2}\gamma_{3}}{\sqrt{2}}\frac{1 + \gamma_{4}\gamma_{5}}{\sqrt{2}}\frac{1 + \gamma_{6}\gamma_{7}}{\sqrt{2}}\frac{1 + \gamma_{8}\gamma_{1}}{\sqrt{2}}.
\end{align}
In the presence of a charging energy, the effective Hamiltonian for an octagonal island acquires terms due to phase-slip events $\varphi_{\beta}\rightarrow \varphi_{\beta} + 2\pi m$, in addition to the ring-exchange term (\ref{eq:H_Oct}) from single-electron tunneling.  
When the gate charge $n_{g}$ on the octagonal island is an integer, the effective Hamiltonian, including the  contribution from the dominant $\pm 2\pi$ phase slips, takes the form:
\begin{align}
H_{\beta}(n_{g}) &= -\frac{5\delta^{4}}{16t_{\alpha}^{3}}\mathcal{O}^{(2)}_{\beta} + \left(t_{\beta}\hat{W} e^{2\pi i n_{g}} + \mathrm{h.c.}\right)\\
&= - \left[\frac{5\delta^{4}}{16t_{\alpha}^{3}} + \frac{t_{\beta}}{4}\right]\mathcal{O}^{(2)}_{\beta} + t_\beta V_{\beta}(n_{g}) \label{new H}
\end{align}
up to an overall constant. Here,  $t_{\beta}$ is the amplitude for a $2\pi$ phase-slip on octagonal island $\beta$, and 
$V_{\beta}(n_{g})$ is a sum of quartic Majorana operators. Note that $t_{\beta}$ is time-dependent; as a function of time, $t_{\beta}$ increases as the charging energy for the octagonal island $\beta$ 
is slowly turned on during the stabilizer measurement and decreases as the charging energy is slowly turned off after the measurement has been completed.  

\begin{figure}
\includegraphics[trim = 27 17 38 10, clip = true, width=0.45\textwidth, angle = 0.]{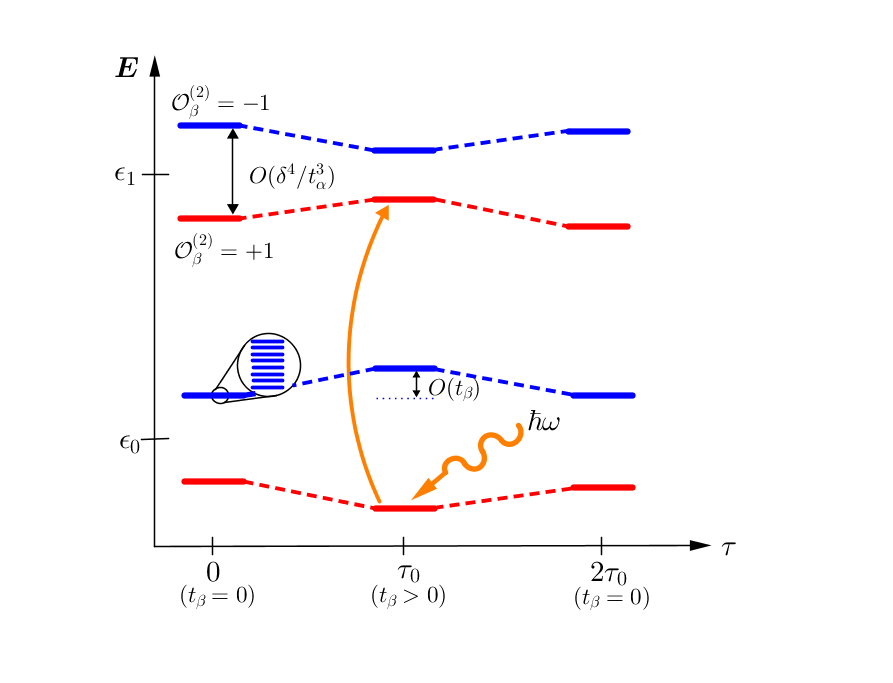}
\caption{We perform an adiabatic measurement of the eight-Majorana stabilizer as follows.  The energy spectrum on an octagonal island originally appears as a series of harmonic oscillator levels ($\epsilon_{n}$), split by the eight-Majorana ring-exchange interaction, as shown on the left.  To measure the stabilizer eigenvalue, we increase the charging energy on an octagonal island to a value $E_{c}$ on a time-scale $\tau \gg t_{\alpha}^{3}/\delta^{4}$.  We now couple the island to photons and measure the energy gap to the next excited state for a sufficiently long duration, so that the gap will differ by $\pm O(t_{\beta})$ depending on the stabilizer eigenvalue $\mathcal{O}_{\beta}^{(2)} = \pm 1$ as shown in the spectrum on the right.  After performing a measurement, we may decrease the charging energy to zero adiabatically, returning to the many-body eigenstate of the ideal stabilizer Hamiltonian with $\mathcal{O}_{\beta}^{(2)} = \pm 1$ as determined by the projective measurement.}\label{fig:Octagon_Measurement}
\end{figure}

As illustrated in Figure \ref{fig:Octagon_Measurement}, we now measure the eight-Majorana stabilizer during the interval $t_\beta \neq 0$ by 
coupling the octagonal island to a resonator and measuring the energy gap to the next harmonic oscillator level.  For the Majorana surface code, the computational states consist of the eigenstates of the  stabilizer Hamiltonian (\ref{eq:Hamiltonian}) that we engineered using an array of topological superconductor islands. The Hamiltonian for the octagonal island is no longer ideal (\ref{new H})  and includes an additional non-commuting term $V_{\beta}$.  However, as the Hamiltonian
is gapped throughout the measurement process and the charging energy was turned on slowly, the many-body state of the system evolves adiabatically.  When $t_{\beta}\ne 0$, a measurement of the energy gap, when performed for a sufficiently long duration, will determine the eigenvalue of $\mathcal{O}_{\beta}^{(2)}$, as in the case of square islands.   After the measurement, we decrease the charging energy on the octagonal island sufficiently slowly, on a time-scale $\tau \gg t_{\alpha}^{3}/\delta^{4}$, so that the many-body state of the system evolves adiabatically back into an energy eigenstate of the ideal stabilizer Hamiltonian with $\mathcal{O}_{\beta}^{(2)} = \pm 1$ as determined by the projective measurement.

We emphasize that this adiabatic measurement protocol is made possible by our Hamiltonian-measurement hybrid approach. The physical Hamiltonian for our square-octagon lattice of Majorana fermions is precisely the ideal Hamiltonian (\ref{eq:Hamiltonian}) whose eigenstates define the set of computational states. Therefore, even if stabilizer measurements are  imperfect, the many-body state of the system will, after a sufficient interval of time, return to the computational basis with the desired stabilizer eigenvalue. This is in contrast to ``measurement-only'' approaches such as the ordinary surface code, where measurement errors cannot be handled in a controlled manner, 
as physical qubits are only coupled by measurements instead of a static interaction Hamiltonian.  



\section{Conclusion}

In summary, we have improved on a recent proposal of implementing a Majorana fermion surface code  in  an array of topological superconductor islands \cite{Maj_Surf_Code}. 
Our proposal combines (1) the engineering a static stabilizer Hamiltonian by using  the charging energy of superconducting islands and electron tunneling between islands with (2) 
single-step projective measurements for quantum error correction and gate operations. Given the rapid experimental progress towards the  
identification of Majorana fermions \cite{delft, weizmann, princeton, jia} and the tantalizing prospect of integration with superconducting qubits \cite{marcus},  
we hope the implementation of a Majorana fermion surface code proposed in this work be pursued.  Recently, we have learned of an attractive proposal on a different implementation of the Majorana fermion surface code 
in an array of semiconductor nanowire based topological superconductors, taking a novel approach to stabilizer measurements \cite{Altland}. 
We also note that lattice systems of interacting Majorana fermion may host a wide variety of intriguing quantum phenomena including symmetry breaking \cite{franz}, 
topological order \cite{teo, sau}, and beyond \cite{vijay}, continuing to inspire new approaches to quantum computation.  
Broadly speaking, the Hamiltonian-measurement hybrid approach introduced in our work may be widely applicable to 
implementing stabilizer codes for large-scale, fault-tolerant quantum computation and is likely to be advantageous over Hamiltonian-only or measurement-only approaches.    

\begin{acknowledgments}
This work was supported by the Packard Foundation (LF), 
and the DOE Office of Basic Energy Sciences, Division of Materials Sciences and Engineering under Award No. DE-SC0010526 (SV).
\end{acknowledgments}

\end{document}